\documentclass[12pt,preprint]{aastex}

\shorttitle{Imaging a solar eruptive flare}
\shortauthors{Li et al.}

\begin{document}

\title{Imaging Observations of Magnetic Reconnection \\in a Solar Eruptive Flare}

\author{Y. Li$^{1,2,3}$, X. Sun$^{4}$, M. D. Ding$^{1,3}$, J. Qiu$^{2}$ \& E. R. Priest$^{5}$}
\affil{$^1$School of Astronomy and Space Science, Nanjing University, Nanjing 210023, China; yingli@nju.edu.cn}
\affil{$^2$Department of Physics, Montana State University, Bozeman, MT 59717, USA}
\affil{$^3$Key Laboratory for Modern Astronomy and Astrophysics (Nanjing University), Ministry of Education, Nanjing 210023, China}
\affil{$^4$W. W. Hansen Experimental Physics Laboratory, Stanford University, Stanford, CA 94305, USA}
\affil{$^5$School of Mathematics and Statistics, University of St Andrews, Fife KY16 9SS, Scotland, UK}

\begin{abstract}
Solar flares are one of the most energetic events in the solar atmosphere. It is widely accepted that flares are powered by magnetic reconnection in the corona. An eruptive flare is usually accompanied by a coronal mass ejection, both of which are probably driven by the eruption of a magnetic flux rope (MFR). Here we report an eruptive flare on 2016 March 23 observed by the Atmospheric Imaging Assembly on board the {\em Solar Dynamics Observatory}. The extreme-ultraviolet imaging observations exhibit the clear rise and eruption of an MFR. In particular, the observations reveal solid evidence for magnetic reconnection from both the corona and chromosphere during the flare. Moreover, weak reconnection is observed before the start of the flare. We find that the preflare weak reconnection is of tether-cutting type and helps the MFR to rise slowly. Induced by a further rise of the MFR, strong reconnection occurs in the rise phases of the flare, which is temporally related to the MFR eruption. We also find that the magnetic reconnection is more of 3D-type in the early phase, as manifested in a strong-to-weak shear transition in flare loops, and becomes more 2D-like in the later phase, as shown by the apparent rising motion of an arcade of flare loops.
\end{abstract}


%--------------------------------------------------------
\section{Introduction}

Solar flares refer to a sudden brightening in the solar atmosphere as a result of a vast magnetic energy release in the corona \citep{shib11,flet11}. An eruptive flare is accompanied by a coronal mass ejection (CME) (e.g., \citealt{schm15}), both of which can play an important role in space weather. There had been a long debate about whether a flare or CME is the primary cause of the whole eruption \citep{kahl92,harr95}. Recently, a consensus has been achieved that they represent two different aspects of one eruptive magnetic process \citep{zhan01,webb12}.

The most popular picture for eruptive flares is known as the standard (or CSHKP; \citealt{carm64,stur66,hira74,kopp76}) model, which may be described as follows. A magnetic flux rope (MFR) rises due to some process of instability or non-equilibrium and stretches the arcade field lines straddling over it. The two legs (with opposite magnetic polarities) of the arcade field approach one another and form a current sheet in between, where magnetic reconnection occurs and releases magnetic energy that was stored beforehand. As a feedback, magnetic reconnection accelerates the overlying MFR to form a CME. It also heats the reconnected loops below, i.e., flare loops, whose footpoints form two elongated flare ribbons in the chromosphere residing on the two sides of a polarity inversion line (PIL). In the two-dimensional (2D) scenario (e.g., \citealt{carg83,tsun97}), magnetic reconnection takes place at an X-type null point, resulting in flare loops at successively higher altitudes and two flare ribbons separating from each other. The MFR is manifested as a plasmoid (or bubble-like) structure. In the realistic 3D case (reviewed by \citealt{prie14}), magnetic reconnection takes place at either a null point \citep{pont13} or a separator \citep{prie96,parn10} or a quasi-separatrix layer (QSL; \citealt{prie95,aula12}). The flare loops display apparent zipping or shearing motions along a separatrix or quasi-separatrix surface \citep{prie92,aula07,janv13}, whose footpoints move along two sheared flare ribbons \citep{aula12,dudi14,prie16}. The MFR comprises a set of twisted field lines whose ends are anchored on the solar surface \citep{prie00,aula10}. In the past, a large number of observations have been presented that lend support to the above model, either in 2D or 3D, including the existence of an MFR \citep{gibs04,gibs06b,gibs08,chen11,zhan12}, evidence for magnetic reconnection \citep{tsun92,shib99,yoko01,suya13,sunj15}, and evolution of flare ribbons \citep{flet01,qiuj10}. However, those observations captured only part of the whole picture owing to the limited sensitivity and resolution of the observing instruments as well as non-optimal viewing angles.

Here, we present extreme-ultraviolet (EUV) imaging observations of an eruptive flare that exhibits many key features depicted in the standard flare model. The observations were obtained by the Atmospheric Imaging Assembly (AIA; \citealt{leme12}) on board the {\em Solar Dynamics Observatory} ({\em SDO}) with a high spatial resolution ($0.\!\!^{\prime\prime}6$ pixel$^{-1}$) and time cadence (12 s). This event, observed at an optimal viewing angle, exhibits the rise and eruption of a bubble-like MFR, and in particular, it clearly reveals evidence of magnetic reconnection from both the corona and chromosphere during the flare. Moreover, weak reconnection is observed before the start of the flare. The observations provide a comprehensive picture of a typical solar eruptive flare with high clarity and highlight the key processes and their mutual relationships involved in the flare.

%--------------------------------------------------------
\section{Overview of the eruptive event}

\begin{figure*}
\centering
\includegraphics[width=9.5cm]{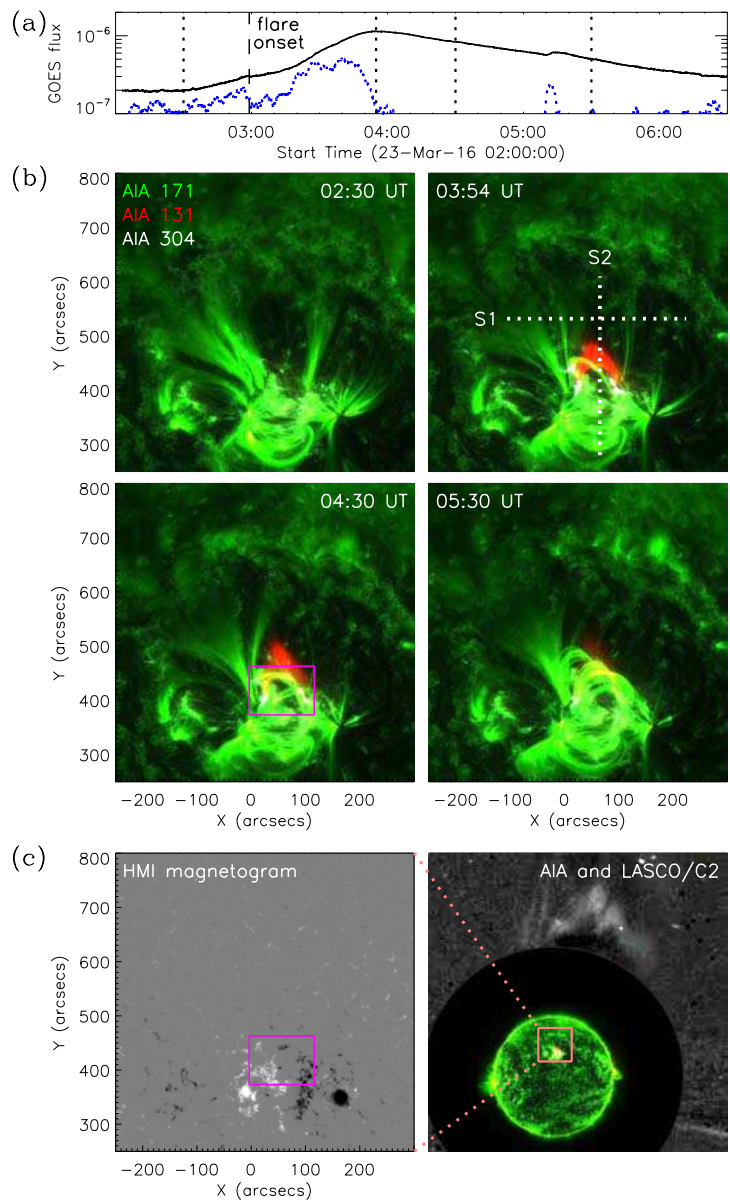}
\caption{{\small Overview of the eruptive event. (a) {\em GOES} 1--8 \AA~soft X-ray flux (black) and its derivative (blue). The vertical dashed line marks the flare onset time (02:59 UT). (b) Composite images of AIA 171 \AA, 131 \AA, and 304 \AA~at four times indicated by the vertical dotted lines in panel (a). Two white dotted lines denote two slices (S1 and S2) that are used to trace different motions shown in Figures \ref{fig-seq}(a) and \ref{fig-pfl}(c). (c) Preflare line of sight HMI magnetogram (left) and composite image of AIA and LASCO C2 (right). The magenta box on the magnetogram (the same as in panel (b)) marks the flare ribbon region shown in Figure \ref{fig-pfl}. The pink box in the composite image shows the whole flaring region.}}
\label{fig-obs}
\end{figure*}

\begin{figure*}
\centering
\includegraphics[width=12cm]{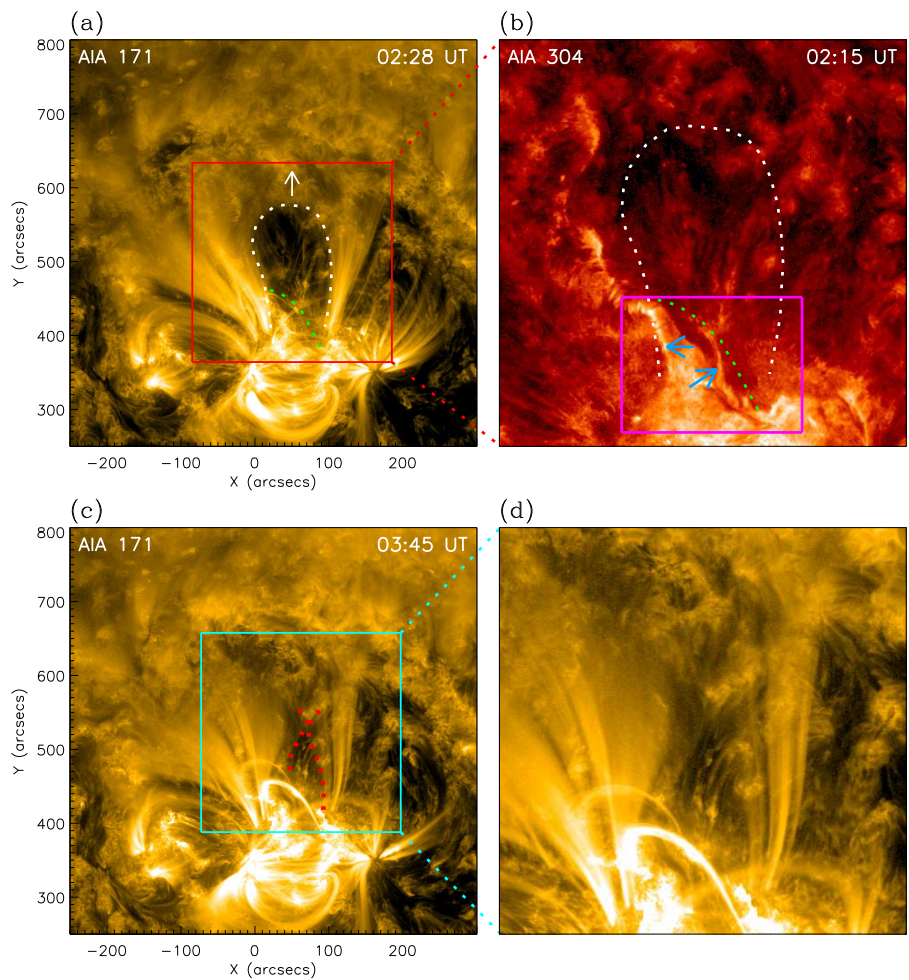}
\caption{{\small Signatures of magnetic reconnection in different phases of the flare. (a) An AIA 171 \AA~image during the preflare phase showing the bubble-like MFR (marked by the white dotted curve). The white arrow indicates the rising of the MFR. The green dotted line traces a highly sheared arch. The red box denotes the field of view of panel (b). (b) An AIA 304 \AA~image during the preflare phase showing two faint ribbons (indicated by the two blue arrows). The magenta box denotes the ribbon region with the same field of view as the magenta box marked in Figure \ref{fig-obs}(c). The white and green dotted curves are the same as in panel (a). (c) An AIA 171 \AA~image in the rise phase of the flare showing the X-type structure (marked by the red dotted curves). The cyan box denotes the field of view of panel (d). (d) Zoom on the X-type structure.}}
\label{fig-rec}
\end{figure*}

The eruptive event here is a C1.1 flare that occurred on 2016 March 23 in active region NOAA 12524 near the solar disk center (N20W04). Based on the {\em GOES} 1--8 \AA~soft X-ray flux (the black curve in Figure \ref{fig-obs}(a)), the flare started at 02:59 UT\footnote{The start time of an X-ray event (i.e., a flare) is defined as the first minute, in a sequence of four minutes, of relatively steep monotonic increase in the {\em GOES} 1--8 \AA~flux (see the blue curve for the derivative of the flux). It is calculated automatically and usually considered as the beginning of the rise phase of the flare.}, peaked at 03:54 UT, and extended into 06:30 UT. Thus, we define the time of 02:59 UT as the flare onset time (see the next Section for more details). Before the onset, we notice that some faint chromospheric ribbons have appeared in the AIA 304 \AA~images (sensitive to a temperature of $\sim$0.05 MK; see Figure \ref{fig-rec}(b)). In the meantime, a pre-existing bubble-like structure, interpreted as an MFR (see the following text), starts to rise (Figures \ref{fig-obs}(b) and \ref{fig-rec}(a) and Animation 1). The rising MFR stretches the overlying coronal loops whose legs move closer to each other towards an X-type structure (Figures \ref{fig-rec}(c) and (d) and also Animation 1). Then a much more rapid reconnection is triggered, at the flare onset, with the plasma near the reconnection site being rapidly heated to $\sim$10 MK and becoming clearly visible in the AIA 131 \AA~passband (the red source in Figure \ref{fig-obs}(b) and Animation 1). Below the X-type structure, flare loops are formed, as clearly shown in the AIA 171 \AA~images ($\sim$0.6 MK; see Figure \ref{fig-obs}(b) and Animation 1). Their footpoints, i.e., two bright flare ribbons, are visible in the AIA 304 \AA~images (the white ribbons in Figure \ref{fig-obs}(b), and also see Animation 1). Above the X-type structure, the MFR is gradually accelerated to form a weak CME, as observed by the Large Angle and Spectrometric Coronagraph (LASCO) on board the {\em Solar and Heliospheric Observatory} (see the right panel of Figure \ref{fig-obs}(c)). The eruptive event takes place in a bipolar magnetic field with the flare ribbons sweeping over plage regions with relatively weak magnetic fields (the magenta box in the left panel of Figure \ref{fig-obs}(c)). The flaring region is next to a pair of sunspots that are less involved in the eruption.

%--------------------------------------------------------
\section{Analysis and results}

\subsection{Preflare phase}

We note that, before 02:59 UT (defined as the flare onset), the {\em GOES} 1--8 \AA~flux started to increase slowly (see Figure \ref{fig-obs}(a)). For the weaker behaviour before the onset time, we have regarded as the preflare phase rather than a separate subflare because it occurs in the same part of the active region as the rise and main phases of the flare. An alternative way of regarding the event would be to say that the whole event began before 02:59 UT, for example, at $\sim$02:30 UT, with a weak rise followed by a strong rise. However, we also note that the faint chromospheric ribbons as well as a highly sheared coronal arch were formed and visible even before 02:30 UT, when the {\em GOES} light curve was quite flat at the preflare background level. The faint ribbons evolved very slowly and only became strong after 02:59 UT when the {\em GOES} 1--8 \AA~flux began to rise impulsively. If we did adopt 02:30 UT as the flare start time, our description of the event and possible causes would be just the same. Figures \ref{fig-rec}(a) and (b) show the faint ribbons in the AIA 304 \AA~image at 02:15 UT (see the blue arrows in the figure and also Animation 1) and the highly sheared arch in the AIA 171 \AA~passband a few minutes later (see the green dotted line in the figure and also Animation 1). These suggest that weak magnetic reconnection has started in the preflare phase.

\subsection{Rise and eruption of the MFR}

\begin{figure*}
\centering
\includegraphics[width=13cm]{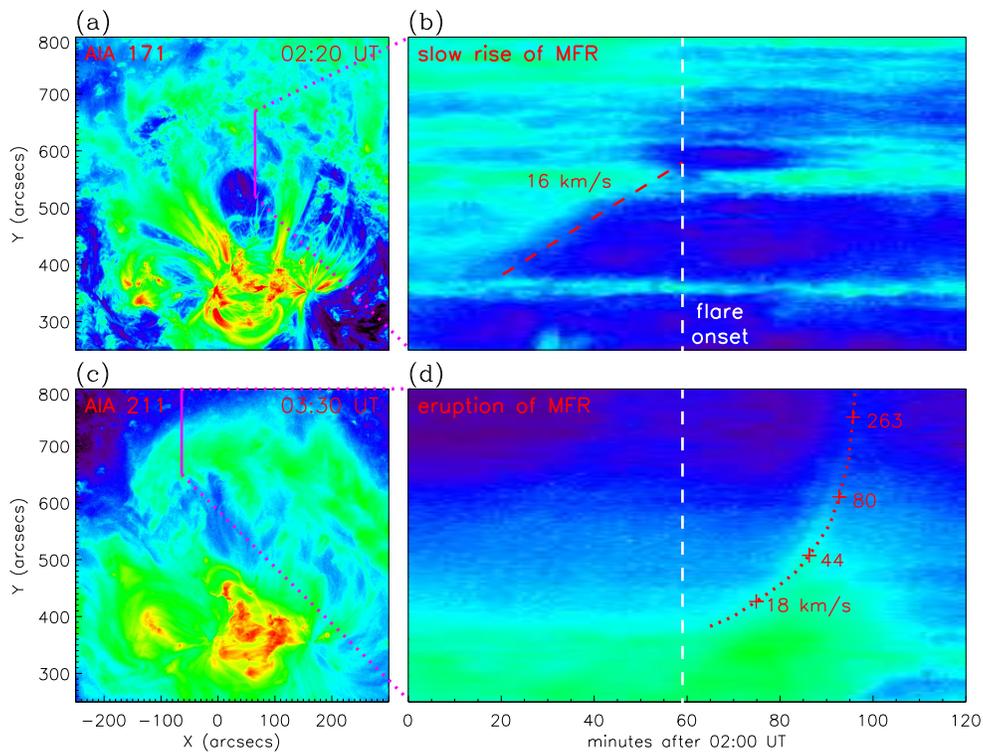}
\caption{{\small Rise and eruption of the MFR. (a) An AIA 171 \AA~image during the preflare phase. The magenta solid line denotes the slice that is used to trace the MFR rise motion. (b) Time-distance map of the slice marked in panel (a). The red dashed line shows the MFR rise motion. The white dashed line denotes the flare onset time (the same as in panel (d)). (c) An AIA 211 \AA~image in the rise phase of the flare. The magenta solid line denotes the slice that is used to trace the MFR eruption. (d) Time-distance map of the slice marked in panel (c). The red dotted line shows the MFR eruptive motion. The speeds are given at several times marked by the plus symbols.}}
\label{fig-mfr}
\end{figure*}

During the preflare phase, a bubble-like structure (or cavity) exists in the corona above the active region as seen in the AIA 171 \AA~passband (indicated by the white dotted curve in Figure \ref{fig-rec}(a)). Such a structure is characteristic of a hot MFR shown in the cool channel when viewed side on \citep{gibs10,chen11,zhan12}. More evidence of the MFR might come from the swirling hooks at the ends of flare ribbons visible in the AIA 171 \AA~and 304 \AA~images (see the cyan boxes in Animation 1 during 02:20--02:50 UT), which could be considered as the MFR footprints \citep{janv15,savc15}. Moreover, some helical structures can be seen in the AIA 193 \AA~and 211 \AA~passbands ($\sim$1.6 MK and $\sim$2.0 MK, respectively; see the black box in Animations 1 and 2 around 03:15 UT). The MFR starts to rise from $\sim$02:20 UT, with an almost constant speed of 16 km s$^{-1}$ initially (Figures \ref{fig-mfr}(a) and (b) and Animation 3). This represents a preflare phase of rise motion for the MFR, which is accompanied by the weak reconnection, as implied by the presence of faint 304 \AA~ribbons and the sheared 171 \AA~arch. After the initial preflare rise of the MFR, the rise phase of the flare begins and we see a much more rapid rise of the MFR, which drives much stronger reconnection to form a CME as observed by LASCO C2. The rapid rise (or eruption) of the MFR is clearly seen in the AIA 211 \AA~passband with a speed from tens to hundreds of km s$^{-1}$ (Figures \ref{fig-mfr}(c) and (d) and Animation 3). The transition from the initial slow rise of the MFR to its rapid eruption is coincident with the flare onset time. 

\subsection{Magnetic reconnection}

\subsubsection{evidence of magnetic reconnection}

\begin{figure*}
\centering
\includegraphics[width=11cm]{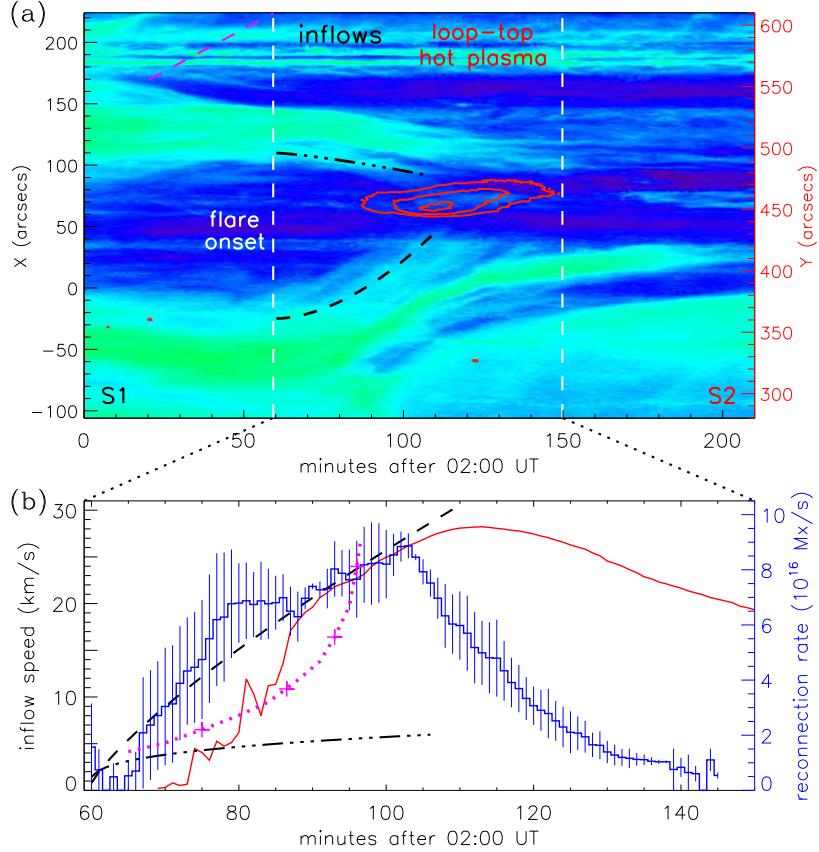}
\caption{{\small Time sequence of key features related to magnetic reconnection. (a) Time-distance map of slices S1 and S2 (marked in Figure \ref{fig-obs}(b)). Two black curves (dashed and dash-dotted) trace the reconnection inflows (corresponding to the left scale along the horizontal slice S1). The red contours denote the loop-top hot source at a temperature of $\sim$10 MK (corresponding to the right scale along the vertical slice S2) and the contour levels represent 13\%, 25\%, and 50\% of the maximum emission at AIA 131 \AA. The magenta dashed line, the same as the red one in Figure \ref{fig-mfr}(b), shows the MFR slow rise motion (corresponding to the slice S2). Two vertical white lines mark the time range in panel (b), with the left one denoting the flare onset. (b) Temporal evolution of reconnection inflow speed (black, left scale), reconnection rate (blue, right scale), and integrated emission at AIA 131 \AA~over the pixels above 13\% of the maximum level (red, arbitrary scale). The magenta dotted curve with plus symbols shows the MFR eruptive motion (an arbitrary scale), the same as the red curve in Figure \ref{fig-mfr}(d).}}
\label{fig-seq}
\end{figure*}

As mentioned above, initial weak reconnection occurred during the preflare phase, at the lower part of the MFR (see the AIA 171 \AA~images at $\sim$02:25 UT in Animation 1). It results in a highly sheared arch (Figure \ref{fig-rec}(a) and Animation 1). Such reconnection is presumably of tether-cutting type \citep{moor92,moor11}, which may help the MFR to rise slowly by reducing the magnetic tension of the overlying field lines \citep{aula10,fany10,schm15}. As the MFR rises, it stretches the overlying arcade field lines whose two legs are pushed closer to form an X-type structure (Figures \ref{fig-rec}(c) and (d)) where rapid flare-related magnetic reconnection commences (i.e., the rise phase of the flare, also see Animation 4). The plasma inflows associated with the reconnection are clearly seen in the time-slice image (tracked by the two black curves in Figure \ref{fig-seq}(a)) with speeds ranging from several to $\sim$30 km s$^{-1}$ (the two black lines in Figure \ref{fig-seq}(b)), which are consistent with previous studies (e.g., \citealt{suya13}). Note that the inflow speeds are different on two sides of the X-type structure, implying an asymmetric magnetic reconnection. The magnetic energy released by reconnection heats the local plasma. The plasma near the reconnection site is heated to at least $\sim$10 MK, as revealed by the greatly enhanced emission at AIA 131 \AA~(the red contours in Figure \ref{fig-seq}(a) and also see Animations 1 and 4). The total flux of this hot passband emission (the red curve in Figure \ref{fig-seq}(b)) appears to be temporally related to the reconnection inflow speeds.
 
Besides the plasma inflows and enhanced hot emission, i.e., evidence for magnetic reconnection in the corona, we observe signatures of reconnection from the chromosphere, namely two brightened flare ribbons as clearly revealed in the AIA 304 \AA~images (Figure \ref{fig-obs}(b) and also Animation 1). Thus, magnetic reconnection rate (time derivative of the reconnection flux) can be measured by adding up the magnetic flux swept out by the newly brightened flare ribbons \citep{forb84,qiuj02}. Here we identify flare ribbon (i.e., footpoint brightening) pixels using the AIA 304 \AA~images. The AIA 304 \AA~emission is enhanced not only in flare ribbons but also in some cooling flare loops, so we only use the data before 05:30 UT when the flare loops start to appear in 304 \AA. Moreover, to distinguish flare ribbon brightenings from other transient non-ribbon features, we define the ribbon brightening pixels to be those whose intensity is enhanced more than 33 times the quiet-Sun level for more than 10 minutes \citep{long07}. Using these thresholds, we trace the flare ribbon brightenings very well for the rise and main phases of the flare. Note that the reconnection rate can be measured in positive and negative magnetic fields, respectively. The imbalance of the positive and negative flux rates could be considered as an uncertainty in the unsigned reconnection rate \citep{flet01,qiuj10}, just like the blue error bar in Figure \ref{fig-seq}(b). It is seen that the reconnection rate (the blue curve in the same figure) grows with the rise of the flare, in particular with the increasing hot emission (the red curve) and the inflow speeds (the two black curves). In addition, the reconnection rate appears to be temporally correlated with the MFR acceleration (the magenta dotted curve, the same as the red curve in Figure \ref{fig-mfr}(d)), indicating a physical link between the flare reconnection and the MFR eruption \citep{qiuj04}. The reconnection rate reaches a maximum of 8.9$\times$10$^{16}$ Mx s$^{-1}$ right before the flare peak time. It then decreases but is still significant 30 minutes later, implying a continuing reconnection throughout the main phase of the flare.

\subsubsection{evolution of magnetic reconnection}

\begin{figure*}
\centering
\includegraphics[width=13cm]{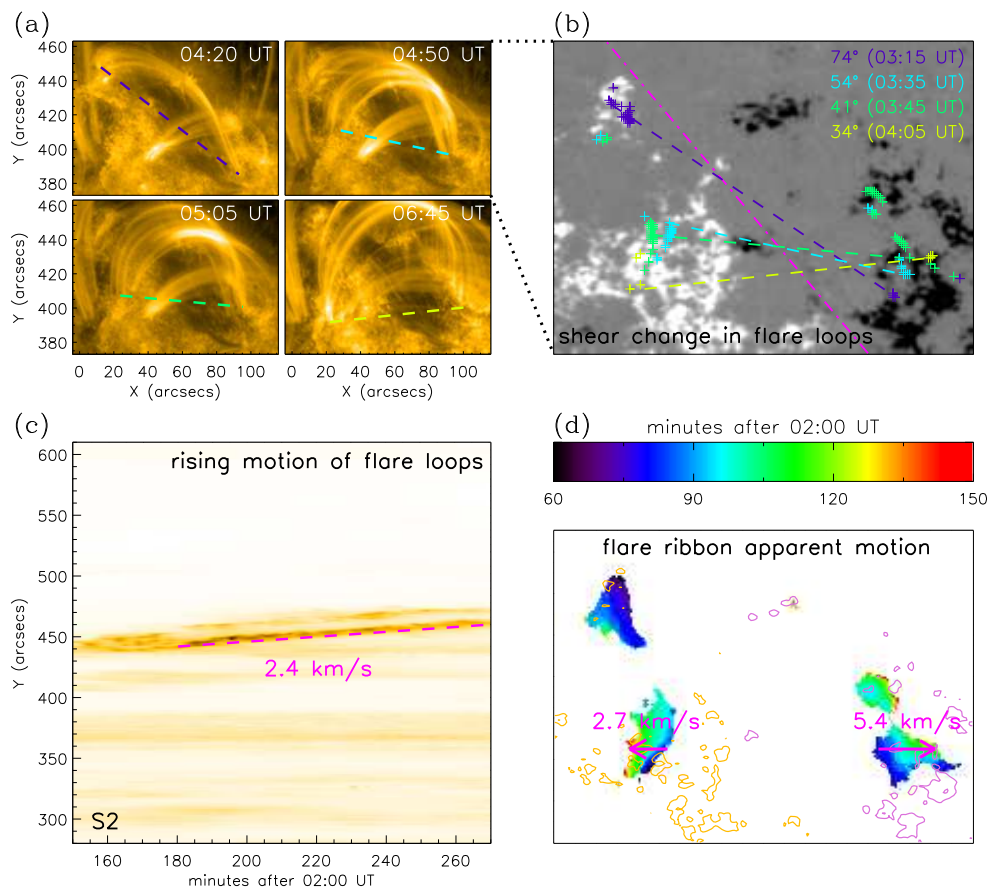}
\caption{{\small Evolution of flare loops and the footpoints. (a) AIA 171 \AA~images at four times in the main phase of the flare with the same field of view as the magenta box marked in Figure \ref{fig-obs}(c). (b) Ribbon brightening pixels (plus symbols) superimposed on the HMI magnetogram. The purple, cyan, green, and yellow dashed lines (also shown in panel (a)) connect a pair of conjugate footpoints brightened at four successive times. The magenta line represents the mean PIL of the flaring region, perpendicular to which is used to measure the shear angle for flare loops. (c) Time-distance map at AIA 171 \AA~along the slice S2 marked in Figure \ref{fig-obs}(b). The dashed line traces the apparent rising motion of flare loops during the main phase. (d) Apparent motion pattern of the flare ribbons in the AIA 304 \AA~images with the same field of view of panel (b). The orange and violet contours mark the magnetic polarities at $+$300 and $-$300 G, respectively. The two magenta arrows are used to measure the average speeds of the ribbon separation.}}
\label{fig-pfl}
\end{figure*}

As an outcome of magnetic reconnection, the evolution of flare loops can reflect how the reconnection evolves. In this event, the AIA 171 \AA~images clearly reveal a strong-to-weak shear variation in the flare loops during the main phase (Figure \ref{fig-pfl}(a) and Animation 1). Accordingly, the footpoints of the loops show apparent shear motions along the chromospheric ribbons (see Figure \ref{fig-pfl}(d)). For a quantitative analysis, we select ribbon brightening pixels at four times, indicated by the plus symbols in purple (03:15 UT), cyan (03:35 UT), green (03:45 UT), and yellow (04:05 UT) (Figure \ref{fig-pfl}(b)), and measure the shear angles at those sample times. Here, the shear angle is defined as the angle between the connecting line of a pair of conjugate footpoints and the line perpendicular to the mean PIL of the flaring region (the magenta line in Figure \ref{fig-pfl}(b)). In this definition, a large shear angle means a strong shear (i.e., more non-potential), while a small angle corresponds to a weak shear (more potential). At the four successive times, the shear angles are measured to be 74$^{\circ}$, 54$^{\circ}$, 41$^{\circ}$, and 34$^{\circ}$, respectively. This indicates that the flare loops are highly sheared early in the main phase but become less sheared at a later time. Such a shear transition in flare loops that are formed at different times and different heights is a typical feature of magnetic reconnection in 3D \citep{suyn06,aula12}.

In the late path of the main phase, we can also see a typical feature of 2D-like magnetic reconnection from less sheared flare loops. As shown in the time-slice image of AIA 171 \AA~(Figure \ref{fig-pfl}(c)), the flare loops exhibit an apparent rising motion with the footpoints separating from each other (Figure \ref{fig-pfl}(d)), which is presumably due to the magnetic reconnection proceeding at higher and higher altitudes. The rise speed, measured in the AIA 171 \AA~images, is $\sim$2.4 km s$^{-1}$. This is comparable with the average ribbon separation speeds measured in the AIA 304 \AA~images, which are $\sim$2.7 and $\sim$5.4 km s$^{-1}$ for the eastern and western footpoint brightenings, respectively (see the two magenta arrows in Figure \ref{fig-pfl}(d)).

%--------------------------------------------------------
\section{Discussions}

Observations of this eruptive flare clearly exhibit many key processes involved in the standard flare model. In particular, the imaging observations show clear evidence for magnetic reconnection from both the corona and chromosphere in different phases of the flare. The EUV images also reveal the rise and eruption of an MFR, which is in close association with the magnetic reconnection.

Consistent with the standard flare model, the EUV observations show solid evidence for strong magnetic reconnection during the rise and main phases of the flare. On the one hand, we detect clear reconnection inflows and cusp-shaped hot emissions in the corona; on the other hand, bright flare ribbons are formed and clearly visible in the chromosphere, and then used to measure the reconnection rate, which is well correlated with the inflow speeds and plasma heating. Note that a number of previous observations have reported evidence of magnetic reconnection in solar flares. For example, recently \cite{suya13} showed inflowing cool loops and outflowing hot loops as well as hot emission in a C2.3 flare. In addition, \cite{sunj15} reported plasma inflows and downflows in a solar eruption. However, the events reported by those authors occur at the solar limb, thus without sufficient information on brightened flare ribbons in the chromosphere. The event in this paper is a nice event on the solar disk to exhibit clear evidence for magnetic reconnection from both the corona and chromosphere with a high level of clarity.

Observed at an optimal viewing angle, this event also provides us a good opportunity to investigate the temporal relationship between the magnetic reconnection and the rising motion of the MFR. We find that faint ribbons and highly sheared arches have appeared in the preflare phase, indicating that weak reconnection has started before the flare onset. Careful inspection shows that the weak reconnection (implied by the appearance of faint ribbons at 02:13 UT $\pm$2 min) is followed by or possibly coincident with the slow rise of the MFR (which starts at 02:20 UT $\pm$5 min). This suggests that the weak reconnection plays at least a partial role in the initial rise of the MFR, or speaking more specifically, it may help the MFR to rise slowly by changing the magnetic connectivity during the preflare phase \citep{prie00,aula10,fany10}. As the MFR rises up, it induces the flare-related strong magnetic reconnection, which in turn accelerates the MFR simultaneously. Therefore, we see a close link between the strong reconnection and the MFR eruption during the flare. The reconnection reduces the restraining effect of the field lines lying over the MFR; then the MFR goes into a state of non-equilibrium with a net upward magnetic force that makes it erupt. Here, we do not exclude that magnetohydrodynamic instabilities (like the kink and torus instabilities) might also play a role in the final eruption \citep{amar00,aula10,fany10,chen13,zucc14}.

Overall, the environment and strength of magnetic reconnection as well as its role in the eruptive event appear to be different during different flare phases. In the initial preflare phase, some highly sheared field lines are involved in weak tether-cutting reconnection and so help the MFR to rise slowly. Induced by a further rise of the MFR, strong reconnection occurs during the rise and main phases, which is temporally related to the MFR eruption. Such reconnection results in an apparent upward motion of flare loops with a strong-to-weak shear variation due to the formation of different flare loops with different magnetic shears on top of one another. In the later main phase, the shear of field lines being reconnected becomes weaker, but the continued rising motion of flare loops with separating footpoints suggests that reconnection continues. The reconnection is thus more 3D-type during the early phases and becomes more 2D-like in the later phase. A model for such a transition of reconnection types and its effect on the flux and magnetic helicity of the MFR has recently been proposed \citep{prie16}.

%--------------------------------------------------------
\section{Summary}

In this paper, we have presented clear and comprehensive observations for a solar eruptive flare. The EUV images from {\em SDO}/AIA exhibit the rise and eruption of an MFR, and in particular, reveal solid evidence for magnetic reconnection during the flare. In addition, weak reconnection is observed in the preflare phase. Our main conclusions are summarized below.

\begin{enumerate}

\item Weak magnetic reconnection starts before the flare onset, as implied by the presence of faint ribbons and highly sheared arches. This reconnection is of tether-cutting type and helps the MFR to rise slowly during the preflare phase.

\item In the rise phase of the flare, strong magnetic reconnection occurs with clear evidence detected from both the corona and chromosphere. The reconnection rate measured from chromospheric ribbons grows with the MFR acceleration, shown as increasing coronal inflow speeds and hot emissions. This correlation indicates a physical link between the flare reconnection and the MFR eruption.

\item The magnetic reconnection shows a transition from more 3D-type to quasi-2D with the flare proceeding, which is reflected from the temporal evolution of flare loops and loop footpoints. The flare loops show mostly a strong-to-weak shear variation in the early phase of the flare, while they mainly show an apparent rising motion with a footpoint separation in the later phase.

\end{enumerate}

%*********************************************************************************************************

\acknowledgments
SDO is a mission of NASA's Living With a Star Program. The authors are very grateful to the referee for his/her valuable comments that significantly improved the manuscript. Y.L. and M.D.D. are supported by NSFC under grants 11373023 and 11403011, and by NKBRSF under grant 2014CB744203. X.S. is supported by NASA Contract NAS5-02139 (HMI) to Stanford University. J.Q. is supported by US NSF grant 1460059. E.R.P. is grateful to the Solar Group at MSU for their warm hospitality during his summer visit and providing some travel support.

\bibliographystyle{apj}

\begin{thebibliography}{}
\expandafter\ifx\csname natexlab\endcsname\relax\def\natexlab#1{#1}\fi

\bibitem[Amari et al.(2000)]{amar00}
 Amari, T., Luciani, J.~F., Mikic, Z., \& Linker, J.\ 2000, \apjl, 529, L49
 
\bibitem[Aulanier et al.(2007)]{aula07}
 Aulanier, G., Golub, L., DeLuca, E.~E., et al.\ 2007, Science, 318, 1588

\bibitem[Aulanier et al.(2012)]{aula12}
 Aulanier, G., Janvier, M., \& Schmieder, B.\ 2012, \aap, 543, A110
 
\bibitem[Aulanier et al.(2010)]{aula10}
 Aulanier, G., T{\"o}r{\"o}k, T., D{\'e}moulin, P., \& DeLuca, E.~E.\ 2010, \apj, 708, 314
 
\bibitem[Cargill \& Priest(1983)]{carg83}
 Cargill, P.~J., \& Priest, E.~R.\ 1983, \apj, 266, 383

\bibitem[Carmichael(1964)]{carm64}
 Carmichael, H.\ 1964, NASA Special Publication, 50, 451

\bibitem[Cheng et al.(2013)]{chen13}
 Cheng, X., Zhang, J., Ding, M.~D., et al.\ 2013, \apjl, 769, L25

\bibitem[Cheng et al.(2011)]{chen11}
 Cheng, X., Zhang, J., Liu, Y., \& Ding, M.~D.\ 2011, \apjl, 732, L25

\bibitem[Dud{\'{\i}}k et al.(2014)]{dudi14}
 Dud{\'{\i}}k, J., Janvier, M., Aulanier, G., et al.\ 2014, \apj, 784, 144

\bibitem[Fan(2010)]{fany10}
 Fan, Y.\ 2010, \apj, 719, 728

\bibitem[Fletcher et al.(2011)]{flet11}
 Fletcher, L., Dennis, B.~R., Hudson, H.~S., et al.\ 2011, \ssr, 159, 19

\bibitem[Fletcher \& Hudson(2001)]{flet01}
 Fletcher, L., \& Hudson, H.\ 2001, \solphys, 204, 69
 
\bibitem[Forbes \& Priest(1984)]{forb84}
 Forbes, T.~G., \& Priest, E.~R.\ 1984, in Solar Terrestrial Physics: Present and Future, ed. D. Butler \& K. Papadopoulos (Washington: NASA), 35

\bibitem[Gibson \& Fan(2008)]{gibs08}
 Gibson, S.~E., \& Fan, Y.\ 2008, Journal of Geophysical Research (Space Physics), 113, A09103

\bibitem[Gibson et al.(2004)]{gibs04}
 Gibson, S.~E., Fan, Y., Mandrini, C., Fisher, G., \& Demoulin, P.\ 2004, \apj, 617, 600

\bibitem[Gibson et al.(2006)]{gibs06b}
 Gibson, S.~E., Fan, Y., T{\"o}r{\"o}k, T., \& Kliem, B.\ 2006, \ssr, 124, 131
 
\bibitem[Gibson et al.(2010)]{gibs10}
 Gibson, S.~E., Kucera, T.~A., Rastawicki, D., et al.\ 2010, \apj, 724, 1133

\bibitem[Harrison(1995)]{harr95} 
 Harrison, R.~A.\ 1995, \aap, 304, 585

\bibitem[Hirayama(1974)]{hira74}
 Hirayama, T.\ 1974, \solphys, 34, 323

\bibitem[Janvier et al.(2015)]{janv15}
 Janvier, M., Aulanier, G., \& D{\'e}moulin, P.\ 2015, \solphys, 290, 3425

\bibitem[Janvier et al.(2013)]{janv13}
 Janvier, M., Aulanier, G., Pariat, E., \& D{\'e}moulin, P.\ 2013, \aap, 555, A77

\bibitem[Kahler(1992)]{kahl92}
 Kahler, S.~W.\ 1992, \araa, 30, 113

\bibitem[Kopp \& Pneuman(1976)]{kopp76}
 Kopp, R.~A., \& Pneuman, G.~W.\ 1976, \solphys, 50, 85

\bibitem[Lemen et al.(2012)]{leme12}
 Lemen, J.~R., Title, A.~M., Akin, D.~J., et al.\ 2012, \solphys, 275, 17

\bibitem[Longcope et al.(2007)]{long07}
 Longcope, D. W., Beveridge, C., Qiu, J., et al.\ 2007, \solphys, 244, 45
 
\bibitem[Moore \& Roumeliotis(1992)]{moor92}
 Moore, R.~L., \& Roumeliotis, G.\ 1992, IAU Colloq.~133: Eruptive Solar Flares, 399, 69
 
\bibitem[Moore et al.(2011)]{moor11}
 Moore, R.~L., Sterling, A.~C., Gary, G.~A., Cirtain, J.~W., \& Falconer, D.~A.\ 2011, \ssr, 160, 73

\bibitem[Parnell et al.(2010)]{parn10}
 Parnell, C.~E., Haynes, A.~L., \& Galsgaard, K.\ 2010, Journal of Geophysical Research (Space Physics), 115, A02102

\bibitem[Pontin et al.(2013)]{pont13}
 Pontin, D.~I., Priest, E.~R., \& Galsgaard, K.\ 2013, \apj, 774, 154

\bibitem[Priest(2014)]{prie14}
 Priest, E.~R.\ 2014, Magnetohydrodynamics of the Sun (Cambridge Univ. Press)

\bibitem[Priest \& D{\'e}moulin(1995)]{prie95}
 Priest, E.~R., \& D{\'e}moulin, P.\ 1995, \jgr, 100, 23443

\bibitem[Priest \& Forbes(1992)]{prie92}
 Priest, E.~R., \& Forbes, T.~G.\ 1992, \jgr, 97, 1521
 
\bibitem[Priest \& Forbes(2000)]{prie00}
 Priest, E.~R., \& Forbes, T.~G.\ 2000, Magnetic reconnection: MHD theory and applications (Cambridge Univ. Press)
 
\bibitem[Priest \& Longcope(2016)]{prie16}
 Priest, E.~R., \& Longcope, D.~W.\ 2016, \solphys, submitted

\bibitem[Priest \& Titov(1996)]{prie96}
 Priest, E.~R., \& Titov, V.~S.\ 1996, Proceedings of the Royal Society of London Series A, 354, 2951
 
\bibitem[Qiu et al.(2002)]{qiuj02}
 Qiu, J., Lee, J., Gary, D.~E., \& Wang, H.\ 2002, \apj, 565, 1335
 
\bibitem[Qiu et al.(2010)]{qiuj10}
 Qiu, J., Liu, W., Hill, N., \& Kazachenko, M.\ 2010, \apj, 725, 319
 
\bibitem[Qiu et al.(2004)]{qiuj04}
 Qiu, J., Wang, H., Cheng, C.~Z., \& Gary, D.~E.\ 2004, \apj, 604, 900

\bibitem[Savcheva et al.(2015)]{savc15}
 Savcheva, A., Pariat, E., McKillop, S., et al.\ 2015, \apj, 810, 96
 
\bibitem[Schmieder et al.(2015)]{schm15}
 Schmieder, B., Aulanier, G., \& Vr{\v s}nak, B.\ 2015, \solphys, 290, 3457

\bibitem[Shibata(1999)]{shib99}
 Shibata, K.\ 1999, \apss, 264, 129
 
\bibitem[Shibata \& Magara(2011)]{shib11}
 Shibata, K., \& Magara, T.\ 2011, Living Reviews in Solar Physics, 8, 6

\bibitem[Sturrock(1966)]{stur66}
 Sturrock, P.~A.\ 1966, \nat, 211, 695

\bibitem[Su et al.(2013)]{suya13}
 Su, Y., Veronig, A.~M., Holman, G.~D., et al.\ 2013, Nature Physics, 9, 489
 
\bibitem[Su et al.(2006)]{suyn06}
 Su, Y.~N., Golub, L., van Ballegooijen, A.~A., \& Gros, M.\ 2006, \solphys, 236, 325
 
\bibitem[Sun et al.(2015)]{sunj15}
 Sun, J.~Q., Cheng, X., Ding, M.~D., et al.\ 2015, Nature Communications, 6, 7598

\bibitem[Tsuneta(1997)]{tsun97}
 Tsuneta, S.\ 1997, \apj, 483, 507
 
\bibitem[Tsuneta et al.(1992)]{tsun92}
 Tsuneta, S., Hara, H., Shimizu, T., et al.\ 1992, \pasj, 44, L63

\bibitem[Webb \& Howard(2012)]{webb12}
 Webb, D.~F., \& Howard, T.~A.\ 2012, Living Reviews in Solar Physics, 9, 3

\bibitem[Yokoyama et al.(2001)]{yoko01}
 Yokoyama, T., Akita, K., Morimoto, T., Inoue, K., \& Newmark, J.\ 2001, \apjl, 546, L69

\bibitem[Zhang et al.(2012)]{zhan12}
 Zhang, J., Cheng, X., \& Ding, M.~D.\ 2012, Nature Communications, 3, 747
 
\bibitem[Zhang et al.(2001)]{zhan01}
 Zhang, J., Dere, K.~P., Howard, R.~A., Kundu, M.~R., \& White, S.~M.\ 2001, \apj, 559, 452
 
\bibitem[Zuccarello et al.(2014)]{zucc14}
 Zuccarello, F.~P., Seaton, D.~B., Mierla, M., et al.\ 2014, \apj, 785, 88

\end{thebibliography}

%--------------------------------------------------------

\end{document}